\documentclass[%
 reprint,
 amsmath,amssymb,
 aps,
]{revtex4-1}

\usepackage{graphicx}
\usepackage{dcolumn}
\usepackage{bm}
\usepackage[utf8]{inputenc}
\usepackage[T1]{fontenc}
\usepackage{mathptmx}
\usepackage{etoolbox}

\begin{document}

\title{Analytic model of stable shock-like structures in laser interaction with underdense plasma for identifying of phase and polarization dependent regime of laser wakefield accelerators} 

\author{Junjue Liao$^1$}

\author{Jihoon Kim$^2$}

\author{Gennady Shvets$^2$}

\affiliation{$^1$Zhiyuan College, Shanghai Jiao Tong University, Shanghai 200240, China}

\affiliation{$^2$School of Applied and Engineering Physics, Cornell University, Ithaca, NY 14850, United States of America}

\date{\today}

\begin{abstract}
We present an analytical model describing a stable shock-like structure that is formed when an ultra-intense laser propagates through an underdense plasma. It is shown that such structures exist in a wide range of laser-plasma parameters, with a unique sub-luminal shock front velocity for each parameter. Numerical methods to accurately describe such shock-front is developed. The formalism is applied to describe the parameter space in which the Carrier-Envelope-Phase (CEP) effect under which phase and polarization dependent super-ponderomotive effects becomes significant. The developed formalism will enable quick identification of regimes in which CEP effects become significant, expediting designing of Laser Wakefield Accelerators operating in the superponderomotive regime.
\end{abstract}

\pacs{}
\maketitle 

\section{Introduction}

Laser Wakefield Accelerators (LWFA) can sustain extremely high accelerating gradients, providing a pathway to ultra-compact electron accelerators \cite{lwfa0,lwfa1,lwfa2,lwfa3}. Plasma electrons are pushed out by the radiation pressure of the driving ultra-intense laser, forming a region devoid of electrons called "plasma bubble" propagating with a speed close to speed of light \cite{bubble1,bubble2,bubble3,bubble4}. The acceleration electric field in "bubble" region can reach dozens of GeV$/m$ \cite{gev1,gev2}.

While the phase-averaged (ponderomotive) description of the plasma response to the laser is usually sufficient \cite{pond1}, phase and polarization dependent Carrier-Envelope Phase (CEP) effect can manifest when the laser self-steepens, increasing to peak amplitude over the time-scale of a single laser cycle \cite{cep1,cep2}. Under such circumstance, the plasma bubble undergoes asymmetric phase and polarization dependent periodic undulation \cite{phase1,phase2,phase3}. This enables control of electron injection and acceleration process giving rise to ultra-high current spatiotemporally structured electron beams \cite{phasehigh1,phase1}.

While several studies have examined the steepened structure and proposed analytical solutions for limited circumstances \cite{compression1,lwfa3,former2,stationary1}, the laser-plasma parameters under which the CEP effect becomes important is still unclear. Namely, a way to estimate the strength of the CEP effect from a given laser-plasma is missing. Such estimation is crucial to identify the laser parameters amenable to observe CEP effects in laser wakefields and designing experiments.

In this paper, we identify the existence of stable shock-like structure using an 1D Quasistatic code Wake1D under a wide range of laser-plasma parameters \cite{pic1}. Based on this finding, we derive an analytical expression that in turn describe the significance of CEP effect for a given laser-plasma parameter. The paper is organized as follows: in section II, we develop a theoretical model for a stable shock-like laser structure based on quasistatic model of laser-plasma interaction \cite{quasi1,former2}, and observations from a quasi-static particle-in-cell simulation WAKE1D. In section III, we analyze the steepening parameters using normalized expressions from this model, which is in turn used to parametrize the importance of CEP for a given laser-plasma parameters.

\section{Persistence of a stable shock front structure}

\begin{figure}[htbp] 
    \centering 
    \includegraphics[width=0.4\textwidth]{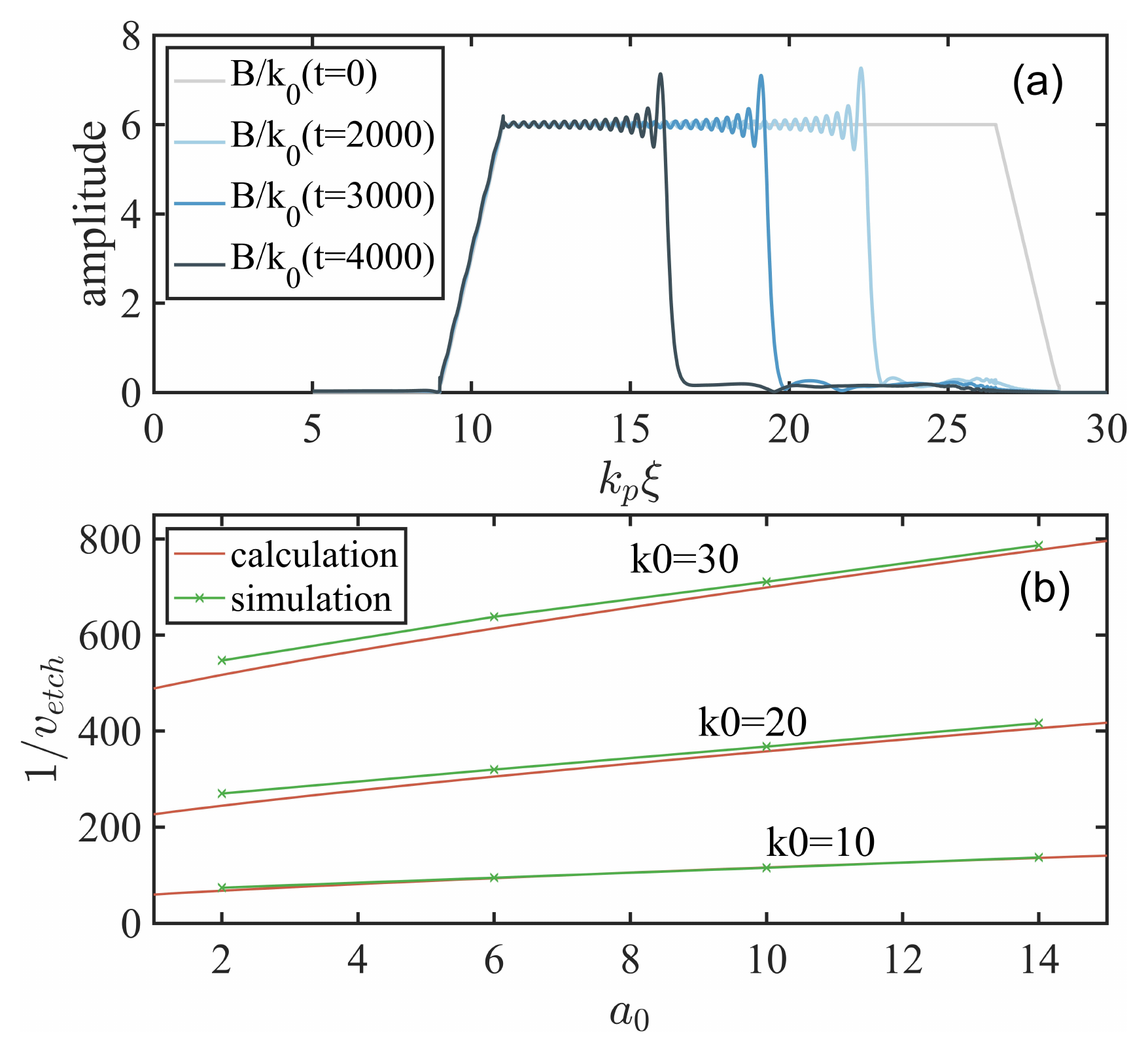} 
    \caption{ Description of the shock front structure in laser wakefield accelerators (a) Formation of the shock-like structure field of flat top laser due to plasma etching in WAKE1D. Simulation parameters: $k_0=20$, $A_0=6$, ramp length $L_{ramp}=2$ (b) Comparison of simulation and theory: Red line is the relation of $a_0$, $k_0$ and $v_{etch}$ obtained by calculation of equation; Green line is the relation of $a_0$, $k_0$ and $v_{etch}$ got by series of PIC simulation} 
    \label{Fig.1} 
\end{figure}

We set the stage by presenting some example results from quasistatic code WAKE1D. The evolution of the steepening front of a Gaussian laser pulse during the etching process is a complex and dynamic phenomenon. To simplify matters, we consider the interaction of a flat-top laser pulse interacting with an underdense plasma.  From a series of PIC simulations covering large range of laser plasma parameters, we observe that there exists a long-lasting stable shock-like structure: a dramatic drop of perpendicular field at the front and oscillation around the initial value behind the front with decreasing disturbance amplitude \cite{stationary1} [Fig. \ref{Fig.1}]. The envelope of pulses used in simulations contain a flat top and ramp at both sides. The length of ramp have no impact on the front structure after complete erosion of ramp component. Based on this observation, we formulate an analytic theory of such shock-like structure using the quasi-static laser plasma formalism.

The quasi-static formalism we use assumes a laser envelope that exhibits slow variations for the laser pulse in comparison to the electron transit time. The distribution of physical quantities is represented in the copropagating frame using the variable $\xi=z-ct$ for the sake of simplicity.  Here we use dimensionless units, normalizing time $t$ to $\omega_p^{-1}$, length $z$ to $k_p^{-1}$,  vector potential $A_0$ to $(e/mc^2)^{-1}$,  electromagnetic fields $E,B$ to $(mc\omega_p/e)^{-1}$, wave vector $k$ to $k_p$  and electron density $n$ to $n_0$. 

For laser driver with frequency $\omega_0$ and vector potential $\tilde{\boldsymbol{A}}_{\perp}=\hat{\boldsymbol{A}}_{\perp} \exp \left[-i k_0 \xi\right]$, the  laser evolution in an 1-D geometry is given by \cite{deplete1}:

\begin{equation}
\left(i k_0 \frac{2 \partial}{\partial t}-\frac{2 \partial^2}{\partial t \partial \xi}\right) \hat{A}_{\perp}= \chi \hat{A}_{\perp},
\end{equation}
where the $\chi=\frac{1}{\psi+1}$ , represents the plasma density accounting for the relativistic transparency \cite{density1}.

The plasma response to the laser pulse is given by \cite{deplete1}:

\begin{equation}
\frac{\partial^2 \psi }{\partial \xi^2}=\frac{1+|\hat A_\perp|^2}{2(1+\psi)^2}-\frac{1}{2}.
\end{equation}

Inserting $\hat{A}_{\perp}=a\exp(i\theta)$, where $a$ and $\theta$ are both real numbers, into Eq. (2):

\begin{equation}
 \left(i k_0 \frac{2 \partial}{\partial t}-\frac{2 \partial^2}{\partial t \partial \xi}\right) (a\exp(i\theta))= \frac{1}{1+\psi} (a\exp(i\theta)).
\end{equation}

Our primary focus is identification of a stable shock-like structure propagating at a constant sub-luminal velocity. We define the speed of front etching in the copropagating frame as $v_{etch}=1-v_{front}$, where $v_{front}$ represents the speed of the front structure in the laboratory frame. Taking note of the etching velocity found in the 1D simulation, we assume that the vector potential can be written as a function of a single variable $\xi-v_{etch}t$, resulting in the transformation of time and space: $\partial_t a= -v_{etch} \partial_\xi a$ and $\partial_t \theta= -v_{etch} \partial_\xi \theta$. We then convert Eq. (3) into a function with only one independent variable $\xi$:

\begin{align}
& i \left(\partial _\xi a \left(k_0-2 \partial_\xi \theta\right)-a \partial_\xi^2 \theta\right)\\
 &+ a (\partial_\xi \theta^2-k_0 \partial_\xi \theta-\partial_\xi^2 a )= -\frac{1}{2v_{etch}}\frac{1}{1+\psi}a.
\end{align}

Given that $a$ is a real number, it is evident that $\partial _\xi a \left(k_0-2 \partial_\xi \theta\right)-a \partial_\xi^2 \theta=0$. By integrating the formula, we can deduce the relationship between the amplitude part and the phase part as follows: $\text{Log} a=-0.5\text{Log}(k_0-2\partial_\xi \theta)+c_1$. The process introduces an indeterminate coefficient $c_1$ into the equation. Boundary condition states that the vector potential must become zero prior to the front owing to depletion, resulting in a numerical divergence of the expression $\frac{c_1^4}{a^4}$. Consequently, we set $c_1=0$, which is equivalent to $\partial_\xi \theta$ being equal to $\frac{k_0}{2}$. The simplified equations are as follows:

\begin{equation}
    \begin{cases}
        \frac{\partial^2 a }{\partial \xi^2}=(-\frac{k_0^2}{4}+\frac{1}{2v_{etch}}\frac{1}{1+\psi})a,\\
        \frac{\partial^2 \psi }{\partial \xi^2}=\frac{a^2}{2(1+\psi)^2}+\frac{1}{2}(\frac{1}{(1+\psi)^2}-1),\\
        \theta=\frac{k_0}{2}(\xi-v_{etch}t).
     \end{cases}
\end{equation}

Similar equations starting from different  assumptions have been derived previously \cite{stationary1}. For ultra-intense laser, $\psi\gg1$ behind the front, where $\frac{\partial^2 a }{\partial \xi^2}\approx(-\frac{k_0^2}{4})a$, which conforms $\partial_\xi \theta=\frac{k_0}{2}$.

The accuracy of the assumptions we made can be evaluated using the WAKE1D PIC. We note that the configuration is determined by $a_0$ and $k_0$ in PIC simulation while it depends on $a_0$ and $v_{etch}$ in our analytic approach. The two different approaches describing the same structure means that there are only 2 degrees of freedom among the three parameters $a_0$, $k_0$ and $v_{etch}$. Thus, the relation of them can be a index of accuracy ; indeed, reasonable accuracy is found for a wide range of parameters for PIC and our approach [Fig \ref{Fig.1}].

\section{Analytical Expressions Representing the front structure}

\subsection{Parameter-free theory of steepened shock-front}

For ultra-intense laser, $\psi\gg 1$ is positioned far behind the front, where $\frac{\partial^2 a }{\partial \xi^2}\approx-\frac{k_0^2}{4}a$. In another word, vector potential distributes like harmonic oscillation with nearly invariable amplitude far behind the front, which can also be seen from Fig(3). We denote this amplitude as $a_m$, which can be a characteristic quantity of the steepened laser pulse.

\begin{figure}[htbp] 
    \centering 
    \includegraphics[width=0.4\textwidth]{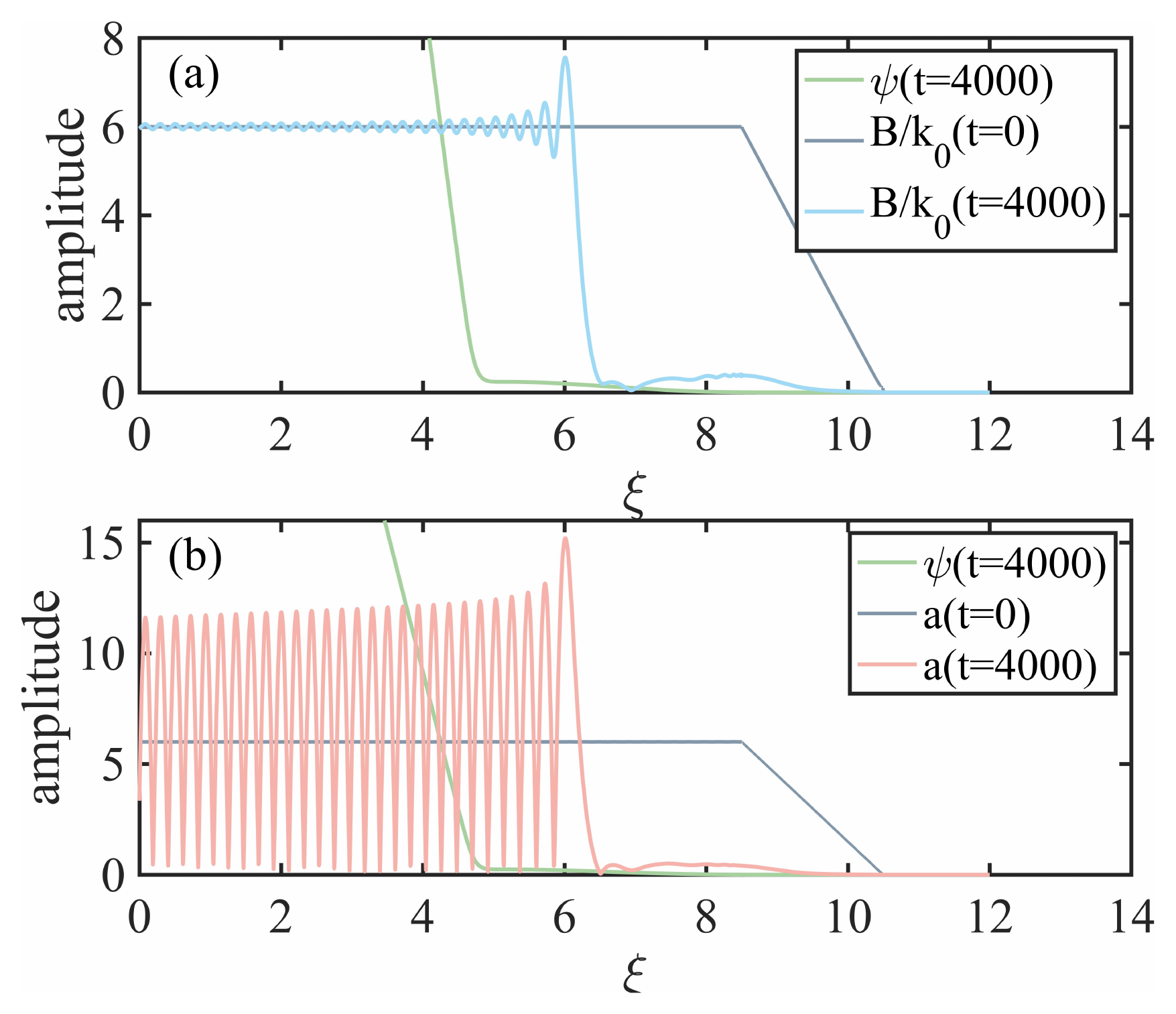} 
    \caption{Distribution of magnetic field (a) and vector potential (b) after the formation of the shock-like structure} 
    \label{Fig.3} 
\end{figure}

For the relation between steepened amplitude $a_{m}$ and initial amplitude $a_0$, the magnetic field needs to be introduced. As for ultra-intense laser, the 'bubble' region behind the front has nearly no electrons to interact with laser. Hence, the magnetic field located at the distant back will maintain its initial value of $a_0k_0$, which can be also observed from Fig(3). The expression of magnetic field is as follows:

\begin{equation}
\tilde{B}_y^L=\frac{\partial \tilde{A}_x}{\partial z}-\frac{\partial \tilde{A}_z}{\partial x} \approx\left(-\frac{\partial \hat{A}_x}{\partial \xi}+i k_0 \hat{A}_x\right) \exp \left(-i k_0 \xi\right).
\end{equation}

Taking advantage of the invariance of the shock-like structure, we use $\hat A_x=\hat{A}_{\perp}=a\exp(i\frac{k_0}{2}(\xi-v_{etch}t))$, to obtain $|\tilde{B}_y^L|^2=\frac{k_0^2}{4}a^2+(\frac{\partial a}{\partial \xi})^2$.

Far behind the front, $\partial_\xi a=0$, and the magnetic field amplitude and vector potential amplitude $a_m$ is given by:

\begin{equation}
|\tilde{B}_y^L|^2=\frac{k_0^2}{4}a_m^2=k_0^2a_0^2,
\end{equation}
showing that the vector potential amplitude behind the shock converges to $a_{m}=2a_0$ and that the relation of $a_0$, $k_0$ and $v_{etch}$ can also be written as $a_0=\frac{a_{m}(k_0,v_{etch})}{2}$. 

From Eq. (6), only periodical solutions exist when $\frac{1}{2v_{etch}}\ge\frac{k_0^2}{4}$ while steepening front solutions exists for opposite case $\frac{1}{2v_{etch}}<\frac{k_0^2}{4}$. This means that when steepening occurs, there is a upper limit of the etching velocity $v_{m}=\frac{2}{k_0^2}$. We define that $v_{etch}=(1-\Gamma)v_m$, where $\Gamma$ is a dimensionless quantity to quantifying the deviation from maximum etching velocity. 

One way to understand this dimensionless quantity $\Gamma$ is using energy balance. The energy depletion rate of laser pulse can be estimated by the etching velocity; namely the amount of energy loss $W$ to the plasma at this shock-like front can be estimated by \cite{energyloss1}:

\begin{equation}
    W=v_{etch}\frac{k_0^2a_0^2}{4\pi}.
\end{equation}

For given $k_0$, the the density of electron interaction with laser at the front is constant. When the vector potential of laser $a_0$ increases, more energy need to be depleted. As a result, the erosion velocity will decrease. Hence, this connection between the energy depletion rate to $v_{etch}$ and $\Gamma$ shows that $\Gamma$ can be considered as a free parameter describing the laser intensity.

Due to the difficulty in obtaining the analytical solution, numerical approaches will be employed to derive the expressions. Extra coefficient in the equation set can be simplified by variable substitution. Assume the normalization units are $\hat a=\frac{\Gamma}{\sqrt{2(1-\Gamma)}}k_0$, $\hat \xi=2 \sqrt{\frac{1-\Gamma}{\Gamma}}\frac{1}{k_0}$. And the normalized variables: $u=a/\hat a$, $f=(1-\frac{1}{1+\psi})/\Gamma$ and $x=\xi/\hat \xi$. Using these variables, Eq. (6) can be recast to:

\begin{equation}
    \begin{cases}
        \frac{\partial^2 u}{\partial x^2}=(1-f)u,\\
        \frac{\partial^2 (1/(\Gamma^2-\Gamma f))} {\partial x^2}=u^2(1-\Gamma f)^2.
     \end{cases}
\end{equation}

As the boundary conditions are known ($u(0)=0$, $\partial_x u|_0=0$, $f(0)=0$,$\partial _x f|_0=0$), the equation set now is only determined by $\Gamma$ instead of two input parameters $k_0$ and $v_{etch}$. Note the nearly stable oscillation amplitude of $u$ behind the front as $u_m$, which is determined by $\Gamma$. In another word, we have:

\begin{equation}
\frac{a_m}{\hat a}=u_m=u_m(\Gamma),
\end{equation}
which can be obtained by one variable fitting.

\begin{equation}
u_m=\frac{2}{\frac{1.19}{\Gamma}-1.46}\frac{\sqrt{2(1-\Gamma)}}{\Gamma}.
\end{equation}

The relative error is within $10\%$  when $\Gamma<0.7$. More accuracy or valid parameter range can be gained by using more sophisticated fitting function. However, $a_0$ is already cover the experiment condition when $\Gamma $ reaches $0.7$, so we retain the expression for the simplicity of form. The real value of $a_0$ can be obtained by $\hat a$ and $u_m$:
\begin{equation}
\begin{aligned}
a_0&=\frac{1}{2}u_m\hat a\\
&=(\frac{1.19}{\Gamma}-1.46)^{-1}k_0.\\
\end{aligned}
\end{equation}

\subsection{Strength of CEP effect using the parameter-free expression}

To observe the CEP effect, the amplitude of the laser potential needs to decrease to zero within one laser wavelength. We refer to this length as the 'width' of the front. To quantify the width, we employ the following definitions: the highest value of the vector potential is denoted as $a_h$, the highest value of $u$ is denoted as $u_h$, the maximum value of the gradient at the front of the vector potential is denoted as $a_{\xi}$, the maximum value of gradient of $u$ is denoted as $u_x$. And the auxiliary variable $\Delta$ is defined as the ratio of $a_h$ to $a_\xi$, serving as a approximate measure of 'width'. Additionally, $\Delta_u$ is defined as $\Delta_u=u_h/u_x$.

From aforementioned analysis, both $u_h$ and $u_x$ are  functions of $\Gamma$:

\begin{align}
&u_h=u_h(\Gamma),\\
&u_x=u_x(\Gamma),\\
&\Delta_u=u_h/u_x=\Delta_u(\Gamma),\\
\end{align}
which can be obtained by fitting.

\begin{equation}
    \Delta_u=\frac{3.6}{\Gamma^{0.51}}\sqrt{\frac{\Gamma}{2(1-\Gamma)}}.
\end{equation}

The relative error is within $5\%$  when $\Gamma<0.9$. The accuracy and valid coverage of parameters is already practical enough. The real value of $\Delta$ be obtained by $\Delta_u$ and normalization units:

\begin{equation}
\begin{aligned}
    \Delta&=\Delta_u \hat \xi\\
    &=\frac{3.56}{\Gamma^{0.5}}\frac{1}{k_0}.\\
\end{aligned}
\end{equation}

As the threshold of CEP being the comparison of $\Delta$ and wavelength, we can have it in parameters space. The $\Delta$ is normalized by $k_p$, so we need $k_0\Delta$ for comparison with laser wavelength.

The result can be verified by the ratio of $k_0\Delta_{theory}$ gained by Eq. (21) and $k_0\Delta_{calculation}$ obtained directly by computing of Eq. (7) minus one. The error $|\frac{k_0\Delta_{theory}}{k_0\Delta_{calculation}}-1|$ is less than $2\%$. Utilising a more advanced fitting primitive can enhance the accuracy but will complicate the formula form.

\begin{figure}[htbp] 
    \centering 
    \includegraphics[width=0.45\textwidth]{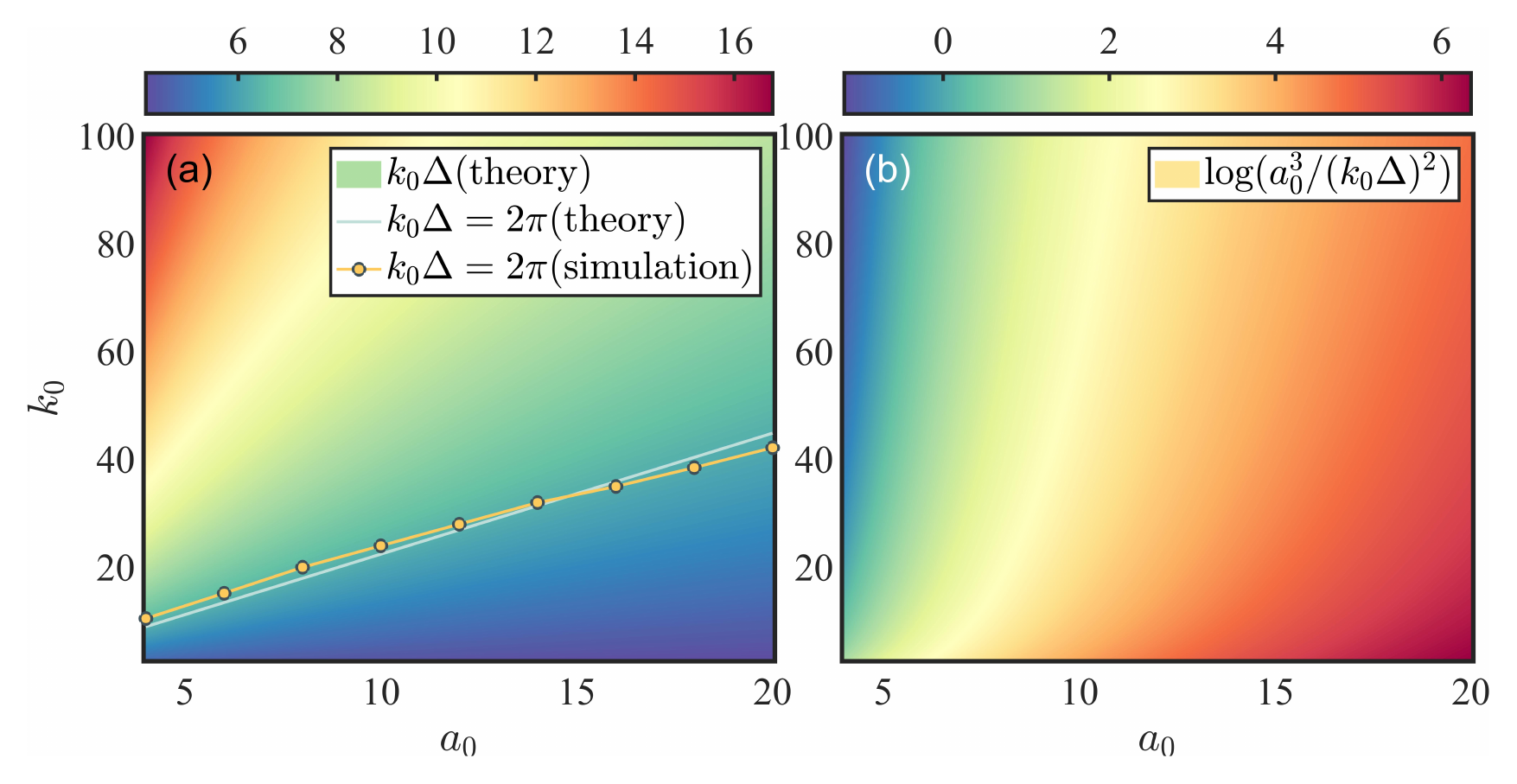} 
    \caption{ (a) $k_0\Delta$ is plotted in the space of $a_0$ and $k_0$. The green line is the theoretical result of  $k_0\Delta=2\pi$, which means the width of front being equal to wavelength of laser. The orange line is the simulation result of $k_0\Delta=2\pi$. (b) $a_0^3/(k_0\Delta)^2$ is plotted in the space of $a_0$ and $k_0$} 
    \label{Fig.5} 
\end{figure}

From previous derivation, both $\frac{a_0}{k_0}$ and $k_0\Delta$ are functions of $\Gamma$:

\begin{align}
    k_0\Delta&=\frac{3.56}{\Gamma^{0.5}},\\
    \frac{k_0}{a_0}&=\frac{1.19}{\Gamma}-1.46.
\end{align}

Thus, the width can be directly determined by laser plasma parameters:

\begin{equation}
    k_0\Delta=3.26(\frac{k_0}{a_0}+1.46)^{0.5}.
\end{equation}

As a result, the contour line of $k_0\Delta$ is a straight line in $a_0$-$k_0$ space, which can be observed in Fig (3). Thus the threshold of CEP effect can be expressed as $a_0 < k_0/2.25$. Although the exact form of the expression Eq. (22) is typically obtained through fitting for practical applications, the functional dependence of $k_0\Delta$ on $\frac{a_0}{k_0}$ is mathematically valid, providing a scaling relation.

In addition, the transverse momenta observed during the CEP effect can serve as another significant indicator: $\Lambda \propto \frac{a_0^3}{\sigma_x^2} \frac{1}{\sigma_z^2} \cos \left(\varphi+\varphi_0\right)$, where $\varphi$ is the CEP and $\varphi_0$ is the CEP offset, $\sigma_x$ and $\sigma_z$ are the transversal and longitudinal spot sizes of a Gaussian Pulse \cite{cep1}. While $\cos(\varphi+\varphi_0)$ illustrates the dependence of electron injection on phase, the ratio $\frac{a_0^3}{\sigma_x^2} \frac{1}{\sigma_z^2}$ quantifies the strength of the coupling between injection and phase. To evaluate the strength in one-dimensional scenario, the term $a_0^3/(k_0\Delta)^2$ in $a_0-k_0$ space is also plotted in Fig (3).

\section{Conclusion}

Recently, study of high-power lasers in relatively dense plasmas have gained a new thrust, as they enable investigation of phase and polarization dependent phenomena in laser wakefield and is capable of generating high-charge and high-current electron beams with spatiotemporal structure. This is due to the CEP effect in which a symmetric laser with a steepened front generates an asymmetric flow. Nevertheless, a predictive model which, given the initial laser intensity and plasma density, can prescribe the strength of the CEP effect has been missing. To remedy this, we developed an hybrid of analytic and numerical approach which can be used to accurately describe the shock-like front structure of a laser propagating inside a plasma. This unique approach enables quick prototyping of experiments or simulations. We envision this model will play an important role in designing experiments capitalizing on CEP effects in existing and upcoming Petawatt-class laser facilities.

\section{acknowledgement}

This work was supported by Zhiyuan College, Shanghai Jiao Tong University. The authors thank Min Chen for inspirational discussions.

\bibliographystyle{unsrt}
\bibliography{main}

\end{document}